\documentclass[12pt,a4paper]{article}

\usepackage[dvips]{graphicx}

\begin{document}
\setlength{\oddsidemargin}{0cm}
\setlength{\baselineskip}{6.3mm}

\begin{titlepage}
            %
    \begin{LARGE}
       \vspace{1cm}
       \begin{center}
        { The Virtual Photon Structure Functions \\
          and AdS/QCD Correspondence
                  }
       \end{center}
    \end{LARGE}

   \vspace{5mm}

\begin{center}
           Yutaka Yoshida\footnote{E-mail address:
             yutaka@scphys.kyoto-u.ac.jp} \\
       \vspace{4mm}
                  {\it Department of Physics, Graduate School of Science, Yoshida South} \\
                  {\it Kyoto University, Kyoto 606-8501, Japan
                  } \\
       \vspace{1cm}

 \begin{abstract}
We study the virtual photon structure functions from gauge/string duality.
If the Bjorken variable $x$ is not small, supergravity approximation becomes good in dual 
string theory. We calculate the virtual photon structure functions at large 't Hooft coupling
in a moderate $x$-region and determine $x$-behavior of the structure functions. 
We also show that the Callan-Gross relation $F_L=0$ is satisfied to a good approximation
 in gravity calculation.
\end{abstract}

\end{center}
\end{titlepage}
\vfill

\date{\today}


\section{Introduction}
The structure functions are important objects in QCD. For example, the nucleon structure functions control the cross section of
deep inelastic scattering and they are related to the parton densities  inside the  initial hadrons. 
The nucleon structure functions however are defined by nucleon matrix elements of the electro-magnetic currents, 
so they cannot be calculated  in the perturbative method.
We can only calculate energy scale  evolution by perturbation theory.  

 AdS/CFT correspondence relates  $\mathcal{N}=4$ super Yang-Mills theory at large 't Hooft coupling in four dimensions
 to weakly-coupled string theory in $AdS_5 \times S^5$ \cite{maldacena}.
The authors of \cite{PS} have studied dual gravity description of nucleon structure functions and have calculated nucleon structure functions.
This is called the hard-wall model.  In this model, the AdS space has a cut-off at infrared region and conformal invariance is broken. 
This cut-off scale corresponds to the infrared mass scale of the gauge theory. At small coupling, probe photons scatter off
partons inside hadrons. But at large 't Hooft coupling, the situation is completely different; probe photons scatter off entire 
hadrons and do not destroy hadronic states. Hadronic states are dual to (massless) string states in AdS space and
we can calculate structure functions from supergravity interactions.
 Usually, the leading-twist operators 
have the leading contribution to OPE of currents correlators at small coupling. But at large 't Hooft coupling,  these operators have
large anomalous dimensions \cite{GKP2} and do not dominate in OPE.

  In this article, we consider the virtual photon structure functions and study their property from bulk dynamics in AdS space.
The photon structure functions are defined by the absorptive part of the four quark currents correlator.
Naively, they can be obtained by replacing the initial hadronic states in hadronic tensors by the target photon states.
So their properties are similar to nucleon structure functions; we can apply standard OPE technique and calculate the anomalous dimensions
of twist-two operators and coefficient functions  perturbatively. A crucial difference of the virtual photon 
structure functions from the nucleon structure functions 
 is that one can determine $x$-behavior of the virtual photon structure functions perturbatively. 
But from theoretical view point, it is still interesting to consider QCD objects at strong coupling which 
are observed at small coupling \cite{HM}.

\section{Review of photon structure functions}
\begin{figure}
\begin{center}
\includegraphics[width=7cm,clip]{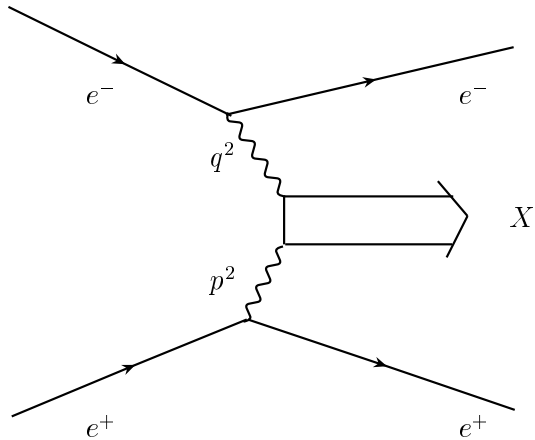}
\caption{$e^- + e^+ \to e^- + e^+ + X$. $X$ represents the final hadron states.  
Wavy lines denote virtual photons which couple to quark currents.
$q^2$ $(p^2)$ is invariant mass squared of the probe (target) photon. }
\label{feyn}
\end{center}
\end{figure}

We review photon structure functions briefly \cite{UW}. 
We take the Lorentz metric $\eta_{\mu \nu }=\rm{diag}(-1,1,1,1) $.
For definiteness, we consider virtual photon scattering on $e^+ e^- \rightarrow e^+ e^- + \rm{hadrons} $ (fig.\ref{feyn}). 
Four dimensional momenta of two virtual photons are $q^{\mu}$ and $p^{\mu}$ $(p^2 \le q^2)$. 
We call the photon whose momentum is $q^{\mu} $ $(p^{\mu})$ the probe (target) photon. 
 If the invariant mass of the target photon  is close to on-shell $p^2 \simeq 0$, 
then vector meson dominance is realized. This process is rather well described by vector meson-photon coupling
than by photon-photon scattering. Dual  gravity descriptions of such interactions are studied in \cite{gomez}, \cite{PRSW}.
If the invariant mass of the target photon  is far off-shell $ \Lambda^2_{QCD} \ll p^2 $, this process is described 
by virtual photon-virtual photon scattering.

We define the tensor $T^{\mu \nu \alpha \beta}$ as
\begin{eqnarray}
T^{ \mu \nu \alpha  \beta} (p,q)=i \int d^4x d^4y d^4z e^{i q\cdot x } e^{i p\cdot (y-z) }
\langle 0 |T( J^{\alpha}(y) J^{\mu}(x)  J^{\nu}(0) J^{\beta}(z) ) | 0 \rangle ,
\end{eqnarray}
where the $J^{\mu}$ are quark currents which couple to the photons. 
The structure tensor of virtual photons is defined by the absorptive part of $T^{\mu \nu \alpha \beta}$.

\begin{eqnarray}
W^{ \mu \nu \alpha  \beta} (p,q)&=&\frac{1}{\pi} {\rm{Im}} T^{\mu \nu \alpha  \beta} (p,q) .
\end{eqnarray} 
This tensor has eight independent components.
If the initial photon states are unpolarized, 
we average the target photon helicities and we obtain
\begin{eqnarray} \label{structure}
W^{\mu \nu } (p,q)&=&\frac{1}{2} \sum_{\lambda} {\epsilon^{(\lambda)}_{\alpha}}^{*} (p) 
W^{ \mu \nu \alpha \beta} (p,q) \epsilon^{(\lambda)}_{\beta} (p)  \nonumber \\
&=&\frac{1}{2} \eta_{\alpha \beta} W^{ \mu \nu \alpha  \beta} (p,q) .
\end{eqnarray}

$W^{\mu \nu }$ can be decomposed into two structure functions $F_1(x,q^2,p^2)$ and $F_{2}(x,q^2,p^2)$ as
\begin{eqnarray}
W^{\mu \nu } (p,q)= ({\eta}^{\mu \nu}-\frac{q^{\mu} q^{\nu}}{q^2}) F_{1}(x,q^2,p^2) 
+(p^{\mu}+\frac{q^{\mu}}{2x})(p^{\nu}+\frac{q^{\nu}}{2x}) \frac{4x}{q^2}F_2 (x,q^2,p^2) , 
\end{eqnarray}
where $x=\frac{-q^2}{2p\cdot q}$. $x$ take $0 \le x \le 1/(1+\frac{p^2}{q^2})$ .

 The structure functions $F_{i}(x,q^2,p^2)$ are called the real photon structure functions in the region $p^2 \simeq 0 $
and  are called the virtual photon structure functions in the region $\Lambda^2_{QCD} \ll p^2 $.

In order to apply OPE for the product of the quark currents which couple to probe photons, the condition $p^2 \ll q^2$ is required.
This means that the distance between quark currents correlators which couple to the probe photons  is much shorter than the others.     
So in the region $\Lambda^2_{QCD} \ll p^2 \ll q^2$, we can calculate the virtual photon structure functions perturbatively at small
coupling.

Inserting the final hadronic states $\sum_{X} | X \rangle \langle X |=1$, we can represent the structure tensor $W^{\mu \nu}$ as
\begin{eqnarray} \label{tens}
W^{\mu \nu} (p,q)&=&\frac{1}{2}
 \langle 0 |\tilde{J}^{\alpha}(-p) \tilde{J}^{\mu}(q)   | X \rangle \langle X |J^{\nu}(0) \tilde{J}_{\alpha}(p)  | 0\rangle \nonumber \\
&=&\frac{1}{2} (2\pi)^4 \sum_{X} \delta^{(4)}(p+q-P_{X})
 \langle 0 |\tilde{J}^{\alpha}(-p)  {J}^{\mu}(0) | X \rangle \langle X |J^{\nu}(0) \tilde{J}_{\alpha}(p)  | 0\rangle , \nonumber \\
\end{eqnarray} 
where we denote $P_{X}$ the momentum of the final state hadron $X$ and $\tilde{J}^{\mu}$ is the Fourier transformation of ${J}^{\mu}$.

\section{Virtual photon structure functions in AdS/QCD}
In this section, we  calculate the virtual photon structure functions from supergravity interaction.
Scattering process is caused by gauged $U(1)$ R-symmetry currents $J^{\mu}$. 
We treat  the process  to the lowest order in $U(1)$ coupling constant and to all orders in strong coupling constant.  
Indices $m, n, \cdots $ denote the $\rm{AdS}_{5}$ space and $\mu, \nu, \cdots $ denote four dimensional Minkowski space. 
Indices $m, n, \cdots $ are raised with the curved metric $g^{m n}$ and $\mu, \nu, \cdots $ are raised with $\eta^{\mu \nu}$.
We define $q^2 = \eta_{\mu \nu} q^{\mu} q^{\nu}$ and $q=\sqrt{q^2}$.
In the hard-wall model, the AdS space has a infrared cut-off at $r_0=\Lambda R^2$ and
$\Lambda$  corresponds to the infrared mass scale of the gauge theory.

The metric of $AdS_5 \times W$ is
\begin{eqnarray}
ds^2=\frac{r^2}{R^2} \eta_{\mu \nu} dy^{\mu} dy^{\nu}+ \frac{R^2}{r^2} d^2 r + R^2 ds^2_{W} 
\end{eqnarray}
where $R=(4\pi g_2 N)^{1/4} \alpha'$ is the AdS radius and $W$ is the five-dimensional internal space which has
a $U(1)$-isometry.
The final state hadron is dual to a string state in ten-dimensional space \cite{PS}. 
We consider this string state is given by the dilaton. 
The dilaton wave function $\Phi_X$ is given by
\begin{eqnarray} \label{dila}
\Phi_X=c_X {P_X}^{1/2} \Lambda^{1/2} \frac{e^{-i P_X \cdot y}}{r^2} J_{\Delta-2} (P_X R^2/r) Y(\Omega),  
\end{eqnarray}
where $c_X$ is a dimensionless constant. $J_{\Delta-2}$ is a Bessel function and $\Delta$ is related to
the  dilaton mass $M^2=\Delta(\Delta-4)/R^2$ in $AdS_5$, and  
$Y(\Omega)$ is the wave function of dilaton in $W$ direction which is normalized

\begin{eqnarray}
\int d^5\Omega \sqrt{g_{W}} |Y(\Omega)|^2 =1 .
\end{eqnarray}

The Kaluza-Klein gauge field $A_m$ couples to the R-current at AdS boundary \cite{GKP1}, \cite{witten2}.
The boundary condition of gauge field is imposed by
\begin{eqnarray}
\lim_{r \to \infty} A_{\mu} (y, r) =\epsilon_{\mu} (q) e^{i q\cdot y} 
\end{eqnarray}
where $\epsilon_{\mu}$ denotes the polarization vector of the gauge field in four dimensions.
The bulk gauge field satisfies the Maxwell equation in the $AdS_5$ space. We take Lorentz gauge in five dimensions, 
then the nonnormalizable mode of gauge field are
\begin{eqnarray}
A_{\mu} (y, r)&=&\epsilon_{\mu} (q) e^{i q\cdot y} \frac{qR^2}{r} K_{1} (q R^2/r), \nonumber \\
A_{r} (y, r)&=&i (\epsilon (q) \cdot q) e^{i q\cdot y} \frac{R^4}{r^3} K_{0} (q R^2/r) ,
\end{eqnarray}
 where $K_0, K_1$ are modified Bessel functions of second type.
Field strength tensors of  $A_{\mu}$ and $A_{r}$ are
\begin{eqnarray} 
F_{\mu \nu}(q)&=&i[ q_{\mu} \epsilon_{\nu}(q)- \epsilon_{\mu}(q) q_{\nu}] \frac{qR^2}{r} K_1 (qR^2/r) e^{iq \cdot y}, \nonumber \\
F_{\mu r}(q)&=&[ \epsilon_{\mu}(q) q^2 - q_{\mu} (\epsilon(q) \cdot q)] \frac{R^4}{r^3} K_0 (qR^2/r) e^{iq \cdot y} . \label{field}
\end{eqnarray}

The insertion of R-symmetry currents excites the metric perturbation $\delta g_{m a}=A_{m}(y,r) v_a(\Omega)$ with
Killing vector $v_a(\Omega)$ associated to a U(1)-isometry of $W$.
 The correlation function of  two R-symmetry currents and a scalar corresponds to the supergravity interaction \cite{freedman}
\begin{eqnarray} 
 \epsilon_{\mu} (q)\epsilon_{\nu} (p) \langle X |\tilde{J}^{\mu}(q)  \tilde{J}^{\nu}(p)  | 0\rangle
\sim \int d^{10}x  \sqrt{-g} \Phi_X F_{mn} F^{mn} v^{a} v_{a} . \label{int}
\end{eqnarray}

From eq.(\ref{field}), eq.(\ref{int}) becomes
\begin{eqnarray} \label{corelator}
\int d^{10}x  \sqrt{-g} \Phi_X F_{mn}(q) F^{mn}(p) v^{a} v_{a} &=&(2\pi)^4 \delta^{(4)}(p+q-P_X) c_X C s^{1/4} \Lambda^{1/2} \nonumber \\
&& \times \int dr \frac{r}{R^3}(\frac{R^4}{r^4} F_{\mu \nu} (q) F^{\mu \nu} (p)+2 F_{\mu r} (q) F^{\mu}_{r}(p) ) \nonumber \\
\end{eqnarray}
where $C=\int \sqrt{g_{W}} Y(\Omega) v^{a} v_{a}$. This factor does not affect $x$-behavior of $F_1(x, q^2, p^2)$ and $F_{2}(x, q^2, p^2)$, so we 
do not write it explicitly, then 

\begin{eqnarray}  
 \epsilon_{\mu} (q)\epsilon_{\nu} (p) \langle X |J^{\mu}(0) \tilde{J}^{\nu}(p)  | 0\rangle 
&=& c_X C s^{1/4} \Lambda^{1/2}  
\int dr \frac{r}{R^3}(\frac{R^4}{r^4} F_{\mu \nu} (q) F^{\mu \nu} (p)+2 F_{\mu r} (q) F^{\mu}_{r}(p) ) \nonumber \\
&=&c_X C s^{1/4} \Lambda^{1/2} \Bigr\{ -2 \Bigr[(q \cdot p)(\epsilon(q) \cdot \epsilon(p)) 
-(q\cdot \epsilon(p)) (p\cdot \epsilon(q)) \Bigr] \nonumber \\ 
&& 
\times q p A(x, q^2, p^2)   + 2\Bigr[(\epsilon(q) \cdot \epsilon(p)) q^2 p^2 \nonumber \\
&&-(p\cdot \epsilon(q)) (p\cdot \epsilon(p)) q^2 -(q\cdot \epsilon(q)) (q\cdot \epsilon(p)) p^2 \nonumber \\
&&+(q \cdot p) (q \cdot \epsilon(q)) (p \cdot \epsilon(p)) \Bigr] B(x,q^2,p^2) \Bigr\}  . \label{aaa}
\end{eqnarray}

In the above equation, $A(x, q^2, p^2)$ and $B(x, q^2, p^2)$ are defined by

\begin{eqnarray} \label{AB}
A(x,q^2,p^2)=\int_{0}^{1/ \Lambda} dr \frac{R^5}{r^5} K_1(qR^2/r) K_1(pR^2/r) J_{\Delta-2} (s^{1/2} R^2/r) , \nonumber \\
B(x,q^2,p^2)=\int_{0}^{1/ \Lambda} dr \frac{R^5}{r^5} K_0(qR^2/r) K_0(pR^2/r) J_{\Delta-2} (s^{1/2} R^2/r) ,
\end{eqnarray}
where $s=-(p+q)^2=q^2(1/x-1-p^2/q^2)$.

We have to sum up  the final-state hadrons to get the structure tensor $W^{\mu \nu}$.
In the hard-wall model, the space has a cut-off at $r_0=\Lambda R^2$,
so the zeros of the Bessel function eq.(\ref{dila}) are at $M_n=n\pi \Lambda$ and the on-shell condition of
the dilaton becomes 
 
\begin{eqnarray} \label{zero}
\sum_{n} \delta(M^2_n-s) \sim (\partial M^2_n/\partial n)^{-1} \sim (2\pi s^{1/2} \Lambda)^{-1} .
\end{eqnarray}
 Using eq.(\ref{aaa}) and eq.(\ref{zero}) and taking average  for the polarization 
vectors of the target photon, we get the structure tensor eq.(\ref{tens})
\begin{eqnarray}
 W^{\mu \nu } (p,q)&=&\frac{|c_X|^2 |C|^2}{\pi} \Bigr[ \left({\eta}^{\mu \nu}-\frac{q^{\mu} q^{\nu}}{q^2} \right) p^2 q^6
\Bigr(\frac{A(x,q^2,p^2)}{2x}+\frac{p}{q}B(x,q^2,p^2) \Bigr)^2 \nonumber \\
&&+\left( p^{\mu}+\frac{q^{\mu}}{2x} \right) \left(p^{\nu}+\frac{q^{\nu}}{2x} \right) p^2 q^4 \Bigr( A(x,q^2,p^2)^2-B(x,q^2,p^2)^2 \Bigr) \Bigr].
\end{eqnarray}
Hence  we have the following structure functions

\begin{eqnarray}  \label{str}
F_1 (x,q^2,p^2) &=&\frac{|c_X|^2 |C|^2}{\pi}  p^2 q^6
\Bigr(\frac{A(x,q^2,p^2)}{2x}+\frac{p}{q}B(x,q^2,p^2) \Bigr)^2,  \\
F_2 (x,q^2,p^2)&=&\frac{|c_X|^2 |C|^2}{4\pi x}p^2 q^6 \Bigr( A(x,q^2,p^2)^2-B(x,q^2,p^2)^2 \Bigr) .
\end{eqnarray}
We stress that the condition $p^2 \ll q^2$ is not used in the supergravity calculation of the structure functions. This is quite different
from pQCD analysis in which the condition $p^2 \ll q^2$ is necessary for applying  OPE to the product of quark currents.

We analyze $F_i (x,q^2,p^2), (i=1,2) $ in the kinematical region $p^2 \ll q^2$. $K_i(qR^2/r)$ damps exponentially in small $r $ region. 
So the region which contributes to integrals in $A(x,q^2,p^2)$ and $B(x,q^2,p^2)$ is  $ qR^2 \le r$. 

To begin with, we evaluate $F_1 (x,q^2,p^2)$.  
We show later that the dominant contribution to $F_i (x,q^2,p^2) $ comes from $A(x,q^2,p^2)$ and 
we can neglect the second term in eq.(\ref{str}) for  $p^2 \ll q^2$.
When $p^2 \ll q^2$ is satisfied, we can also use the leading behavior $K_1(pR^2/r) \simeq r/pR^2$.
Then, with an approximation $1/ \Lambda \sim \infty$, this integrals can be performed  and we get

\begin{eqnarray}
A(x,q^2,p^2) &\simeq& \int_{0}^{\infty} dr \frac{R^3}{p r^4} K_1(qR^2/r) J_{\Delta-2} (s^{1/2} R^2/r) \nonumber \\
&=&\frac{2}{p q^3 R^3} \left(\frac{1}{x} - 1 -\frac{p^2}{q^2}\right)^{\frac{\Delta}{2}-1} 
\frac{\Gamma(\frac{\Delta+2}{2})\Gamma(\frac{\Delta}{2})}{\Gamma(\Delta-1)}
\quad  _{2}F_{1}(\frac{\Delta+2}{2},\frac{\Delta}{2};\Delta-1;1-\frac{1}{x} +\frac{p^2}{q^2})  \nonumber \\
&=&\frac{2}{p q^3 R^3} x^2 \Bigr[1 - (1+\frac{p^2}{q^2})x \Bigr]^{\frac{\Delta}{2}-1} \Bigr(1-\frac{p^2}{q^2}x \Bigr)^{1-\frac{\Delta}{2}} 
\frac{\Gamma(\frac{\Delta+2}{2})\Gamma(\frac{\Delta}{2})}{\Gamma(\Delta-1)} \nonumber \\
&& \times {}_2F_{1}(\frac{\Delta+2}{2},\frac{\Delta-2}{2};\Delta-1;\frac{1-(1+\frac{p^2}{q^2})x}{1-\frac{p^2}{q^2}x })  \nonumber \\
\end{eqnarray}
where $ {}_2F_{1} (a,b;c;z)$ is a hypergeometric function.
 From the second line to the third line we use the identity
$ {}_2F_{1} (a,b;c;z)=(1-z)^{-a} {}_2F_{1} (a,c-b;c;z/(z-1))$.
Then $F_{1}$ becomes
\begin{eqnarray} \label{f1}
F_1 (x,q^2,p^2) &\simeq&\frac{|c_X|^2 |C|^2}{4\pi x^2}  p^2 q^6 A(x,q^2,p^2)^2 \nonumber \\
&\simeq&\frac{|c_X|^2 |C|^2 }{\pi R^6}  \frac{\Gamma(\frac{\Delta+2}{2})^2 \Gamma(\frac{\Delta}{2})^2}{\Gamma(\Delta-1)^2}
 x^2 \Bigr[1 - (1+\frac{p^2}{q^2})x \Bigr]^{\frac{\Delta}{2}-1} \Bigr(1-\frac{p^2}{q^2}x \Bigr)^{1-\frac{\Delta}{2}} 
 \nonumber \\
&& \times {}_2F_{1}(\frac{\Delta+2}{2},\frac{\Delta-2}{2};\Delta-1;\frac{1-(1+\frac{p^2}{q^2})x}{1-\frac{p^2}{q^2}x }).  \nonumber \\
\end{eqnarray}
One can see that $F_1 (x,q^2,p^2)$ increases with $x$ in the small $x$-region  and decreases near $x_{\rm{max}} =1/(1+\frac{p^2}{q^2}) $. 
The structure functions should vanish at $x_{\rm{max}}$ and actually  eq.(\ref{f1})  satisfies this requirement.
From eq.(\ref{f1}), we have found that $F_1(x,q^2,p^2) $ scales as $F_1(x)$ in the limit $p^2/q^2 \to 0$.
This is a crucial difference between the photon structure functions and the nucleon structure functions in gravity calculation.       
If we define $\Delta_{i}$ as the conformal dimension of the dilaton or  the dilatino  which is dual to the initial state hadron,
the momentum dependence of  the nucleon structure functions always behave $(\Lambda/q)^{2\Delta_i-2}$ \cite{PS}.
Therefore the nucleon structure functions do not have the scaling property in general $\Delta_{i}$.
The nucleon structure functions have the scaling property  when the conformal dimension is special value $\Delta_{i}=1$ \cite{PRSW}.

Next, we evaluate $F_{2}$.
We transform $w=qR^2/r$ in eq.(\ref{AB}), $B(x,q^2,p^2) $ can be written as
\begin{eqnarray}
\frac{1}{R^3 q^4}\int_{0}^{\infty} dw w^3 K_0(wp/q) K_0(w) J_{\Delta-2} (ws^{1/2}/q).
\end{eqnarray}
When the condition  $p^2 \ll q^2 $ is satisfied, we can use the asymptotic form $K_0(wp/q) \simeq \log (2q/wp)$.
Then
\begin{eqnarray}
B(x,q^2,p^2)&=&\frac{1}{R^3 q^4} \lim_{\epsilon \to 0}\int_{0}^{\infty} 
dw w^{3} \frac{(2q/wp)^{\epsilon}-1}{\epsilon}  K_0(w) J_{\Delta-2} (ws^{1/2}/q). \nonumber \\
&\simeq& \frac{2^2\log(q/p)}{R^3 q^4} \frac{\Gamma(\frac{\Delta+2}{2})^2}{ \Gamma(\Delta-1)}x^2(1-x)^{\frac{\Delta}{2}-1} \nonumber \\
&&\times {}_2F_{1}(\frac{\Delta+2}{2},\frac{\Delta}{2}-2;\Delta-1;1-x) +O(q^{-4}),
\end{eqnarray}
where we used $\log x= \lim_{\epsilon \to 0}(x^{\epsilon}-1)/\epsilon$.
Therefore $B(x,q^2,p^2)^2$ is a subleading  contribution  to $F_{2}$.  
From the expression for $F_2$, we calculate the longitudinal structure function defined as $F_{L}= F_{2}-xF_{1}$.

\begin{eqnarray} 
F_L (x,q^2,p^2)&\simeq&\frac{|c_X|^2 |C|^2}{4\pi x}p^2 q^6 B(x,q^2,p^2)^2 \nonumber \\
&\simeq&\frac{2^2|c_X|^2 |C|^2}{\pi } \frac{\Gamma(\frac{\Delta+2}{2})^4}{ \Gamma(\Delta-1)^2} \frac{p^2}{q^2} \log^2(q/p) \nonumber \\
&& \times x^3(1-x)^{\Delta-2} {}_2F_{1}(\frac{\Delta+2}{2},\frac{\Delta}{2}-2;\Delta-1;1-x)^2 .
\end{eqnarray}

From the above equation, in the limit $p^2 /q^2 \to 0$,
one can see that the Callan-Gross relation $F_{L}=0$ holds in gravity calculation. 
In QCD, the Callan-Gross relation implies the partons involved in the scattering process are spin 1/2 particles.
This relation is weakly broken by higher order corrections in real QCD.
In our calculation, the Callan-Gross relation is weakly broken by the effect of gauge field in radial direction.

\section{Summary and discussions}
In this article, we have calculated the virtual photon structure functions in the hard-wall model. 
The structure functions $F_i (x,q^2,p^2)$ $(i=1,2)$ we obtained possess the following properties which we have in QCD. 
 
1) $F_i$'s increase at small $x$-region, decrease at large $x$-region and vanish at $x=x_{\rm{max}}$.

2)  $F_{L}$ is small compared to $F_{i}$'s. It goes to zero when the limit $p^2 \ll q^2$ as $q^2 \to \infty $. 

It seems peculiar that the Callan-Gross relation holds in gravity calculation, although there exists no spin 1/2 particles in our case.
But as mentioned in \cite{gomez}, the interaction $\Phi F^2$ in eq.(\ref{int}) is 
the low-energy effective action describing a higgs decay into two gammas through a heavy quark triangle 
loop \cite{shifman}. Since the heavy quark is a spin 1/2 particle , this might be a reason why the Callan-Gross relation holds in our calculation.

We simply comment some extension of the virtual photon structure functions at strong coupling.  
In the hard-wall model, mass square of dilatons and radial numbers exhibit quadratic relation, but actually relations mass square
between radial number is linear (Regge trajectory).  A background which reproduces Regge trajectory is 
proposed in \cite{karch}. This is called the soft-wall model. Deep inelastic scattering in the soft-wall model
is studied and the leading order results in soft wall model is same as the one in the hard-wall model \cite{BBNB1}.
We treat the virtual photon structure functions in not so small $x$-region.  
At small $x$-region, we have to take into account excited string states\cite{PS}, \cite{BPST}.
Then, we cannot use the insertion of dilaton states and have to treat four currents correlator directly.
The  analysis would become  much more complicated.  
It is interesting to investigate the virtual photon structure functions in such situations.

\section*{Acknowledgments}
We are grateful to Y.Kitadono, T.Matsuo, K.Sugiyama and T.Uematsu for useful discussions.
We also thanks T.Uematsu for careful reading of the manuscript and  helpful comments.

\end{document}